\documentclass{aa}    
\usepackage{graphicx}
\usepackage{txfonts}
\begin{document}
\title{Low heat conduction in white dwarf boundary layers ?}


\author{F.K. Liu \inst{1}, F. Meyer \inst{2}, E. Meyer-Hofmeister
\inst{2} and V. Burwitz \inst{3}
} 
\offprints{Emmi Meyer-Hofmeister; emm@mpa-garching.mpg.de}
\institute
     {Astronomy Department, Peking University, Beijing 100871, China\\
     \email{fkliu@pku.edu.cn}
     \and Max-Planck-Institut f\"ur Astrophysik, Karl-
     Schwarzschildstr.~1, D-85740 Garching, Germany\\
     \email{frm@mpa-garching.mpg.de;emm@mpa-garching.mpg.de}
     \and {Max-Planck-Institut f\"ur extraterrestrische Physik,
     P.O. Box 1312, D85741 Garching, Germany\\
     \email{burwitz@mpe.mpg.de}
     }}

\date{Received: / Accepted:}

\abstract
 {X-ray spectra of dwarf novae in quiescence observed by Chandra and 
  XMM-Newton provide new information on the boundary layers of their 
  accreting white dwarfs.} 
  {Comparison of observations and models allows us to extract estimates
  for the thermal conductivity in the accretion layer and reach
  conclusions on the relevant physical processes.}
  {We calculate the structure of the dense thermal boundary layer that
  forms under gravity and cooling at the white dwarf surface on
  accretion of gas from a hot tenuous ADAF-type coronal inflow. The 
  distribution of density and temperature obtained allows us to calculate 
  the strength and spectrum of the emitted X-ray radiation. They depend
  strongly on the values of thermal conductivity and mass accretion rate.}
  {We apply our model to the dwarf nova system VW Hyi and compare the
  spectra predicted for different values of the thermal conductivity
  with the observed spectrum. We find a significant deviation for all
  values of thermal conductivity that are a sizable fraction of the
  Spitzer conductivity. A good fit arises however for a conductivity
  of about 1\% of the Spitzer value. This also seems to hold for other
  dwarf nova systems in quiescence. We compare this result with
  thermal conduction in other astrophysical situations.}
 {The highly reduced thermal conductivity in the boundary layer 
  requires magnetic fields perpendicular to the temperature gradient. 
  Locating their origin in the accretion of magnetic fields from the 
  hot ADAF-type coronal flow we find that dynamical 
  effects of these fields will lead to a  spatially intermittent, 
  localized accretion geometry at the white dwarf surface.}

\keywords{ dwarf novae, cataclysmic variables -- accretion -- 
X-rays: individuals: VW Hyi -- conduction -- magnetic fields -- cooling flows}

\maketitle
%
%
\section{Introduction}
Cataclysmic variables (CVs) are interacting binaries in which mass flows 
from a low mass companion star and accretes on the white dwarf
primary. The X-ray emission of the large number of cataclysmic 
variables makes up a large fraction of the apparently diffuse Galactic
Ridge X-ray emission, as was recently found with the help of deep surveys
by \it{Chandra} \rm{and} \it{XMM-Newton} \rm{ (Sazonov et al. 2006). 
In dwarf nova systems, non-magnetic CVs, the accretion onto the white 
dwarf occurs via
an accretion disk and the gas must dissipate its rotational kinetic
energy before it settles on the surface of the more slowly rotating white
dwarf. The X-ray emission is thought to originate from the interface 
between the white dwarf and the inner edge of the disk, a boundary
layer close to the surface. The disk is too cool to contribute
significantly to the X-ray emission.

Different models have been proposed for the structure of the X-ray
emitting region in dwarf nova at low accretion rates, i.e. in their
quiescent state: (1) a cooling flow model based on Mushotzky \&
Szymkowiak (1988), which assumes isobaric radiative cooling of the
gas; (2) a model of a hot boundary layer where heat advection is 
dominant (Narayan \& Popham 1993); (3) a thermal boundary layer model
where heat conduction determines the structure of the spherically
accreting gas around the white dwarf (Liu, Meyer \& Meyer-Hofmeister
1995) and a similar X-ray emitting corona model of Mahasena \& Osaki
(1999); and (4) hot
settling flow solutions ( Narayan \& Medvedev 2001, Medvedev \& Menou
2002) in which viscously mediated losses of rotational energy are an
important parameter. In these papers references to earlier
discussions of emission mechanisms are given.

The much improved sensitivity and spectral resolution of the
new X-ray telescopes allow us to test specific models. The first
tests were performed by Pandel et al. (2003) for
VW Hyi in quiescence and by Perna et al. (2003) for WX Hydri in
quiescence. Pandel et al. (2003) used X-ray and ultraviolet
data obtained with \it{XMM-Newton}\rm{. The authors found that the X-ray
spectrum indicates the presence of optically thin plasma in the
boundary layer that cools as it settles on the white dwarf, and that
the plasma has a range of temperatures that is well described by a
power law or a cooling flow model with a maximum temperature of 6-8
keV. Perna et al. (2003) used }\it{Chandra}\rm{ observations for their 
investigation, computed spectra for the available theoretical
models, i.e. hot boundary layers, hot settling flows and X-ray emitting
coronae. They came to the conclusion that the continuum is reproduced
well by most of the models, but none of them can fully account for the
relative line strengths over the entire spectral range. Pandel et al. 
(2005) studied 10 dwarf novae, based on }\it{XMM-Newton} \rm{ 
observations, including the earlier results for VW Hyi by Pandel et al. (2003).
The authors conclude that the X-ray emission originates from a hot,
optically thin multi-temperature plasma with a temperature
distribution in close agreement with an isobaric cooling flow,
pointing to a cooling plasma settling onto the white dwarf as the
source of X-rays. 

We use the model description worked out earlier (Liu et
al. 1995), which is related to the ``siphon flow model'' for a corona
above an accretion disk in quiescence (Meyer \& Meyer-Hofmeister
1994). In this picture matter in the innermost disk is evaporated into
a coronal flow which provides a geometrically thick gas flow towards
and around the white dwarf. (Dwarf nova systems have cycles of longer
lasting quiescent phases and shorter outbursts. These outbursts are
triggered by a disk instability, then the matter accumulated in the disk
during quiescence accretes with a high mass flow rate
onto the white dwarf (Meyer \& Meyer-Hofmeister 1984). For the
formation of the inner disk hole see Liu et al. (1997)). Since the
flow of gas from the surrounding corona is an essential feature of our
model our approach differs from the simulation of the
boundary layer between a white dwarf and a thin accretion disk which extends
all the way down to the stellar surface, as studied recently by Balsara
et al. (2007). These numerical simulations give insight into the
spreading of the boundary layer, but no conclusions on spectra are derived.

We have chosen the well documented dwarf nova VW Hyi for our new
investigation. We compute the thermal boundary layer structure,
evaluate the emission measures and determine spectra using the XSPEC
package. We come to the somewhat surprising result that only
structures based on very low conductivity give spectra in agreement
with the observations. Pandel et al. (2005) already remarked
 that heat conduction does not dominate over cooling via X-ray
emission because otherwise $T_{\rm{max}}$, the initial temperature
of the cooling gas, would be much smaller than the virial
temperature}$T_{\rm{vir}}$ (the temperature the gas would have if all
the rotational energy from its Keplerian motion were instantly
converted into heat). Indeed high temperatures are needed to obtain a
spectrum such as the  one observed for VW Hyi. The aim of our paper is to
study the structure of the boundary layer under heat conduction. We
discuss what might cause  low heat conduction that seems different from
 heat conduction in other astrophysical situations. As the analysis of
thermal conduction in clusters of galaxies by Narayan and Medvedev
(2001) shows, a low conductivity would point to a non-chaotic magnetic field.

In Sect. 2 we describe the observations we use for our comparison. 
In Sect. 3 we briefly discuss the physics entering our structure
computations as well as the boundary conditions at the bottom
and the top of the boundary layer. We
compare different models in Sect. 4. In Sect. 5. the results of the 
structure computations are presented, in 
Sect. 6 the evaluated spectra. We argue that the spectra of other
dwarf novae also indicate low thermal conductivity in these
 systems. We discuss the presence of low or high thermal conductivity
indicated in different astrophysical sources as clusters of galaxies
and the intercluster medium (Sect. 7). The low conductivity found 
here possibly indicates an important role of the magnetic field
in the boundary layer accretion process. Our conclusions follow in
Sect. 8.

\section{X-ray observations}
For the comparison of our theoretical spectra with observation we use
the \it { XMM-Newton} \rm  spectrum (Jansen 2001) of VW Hyi obtained
with the European Photon Imaging Camera (EPIC)  (Str\"uder et al. 2001).
 This system was observed with the EPIC-pn in 
full-frame mode on 2001 October 19 for 16.1 ks, from
06:10:05 to 10:38:34, in quiescent state, 22 days after a normal
outburst and 23 days before the next outburst (which was a
superoutburst). The spectrum was extracted using SAS version. 
The observed spectrum is shown in Fig. \ref {spectrum}.
The same data for VW Hyi have also been used by Pandel et
al. (2003) and were included in the work by Pandel et al. (2005).

\section{The boundary layer structure}
We describe the accretion process through the boundary layer around
the white dwarf mainly as presented in an earlier work (Liu et al. 1995). 
Passing through a turbulent region, the accreting matter 
totally loses its angular momentum, becomes subsonic and forms a
layer around the white dwarf. We assume that the density is higher
towards the equatorial plane, so that accretion occurs
onto a belt around the white dwarf. We assume that its area is
about half of the full stellar surface area. We take
spherical coordinates with $r$ the distance to the white dwarf center
, $r_*$ the white dwarf radius, and $S=r_*-r$ the height above the
white dwarf surface (negative sign). The downward directed thermal conductive
flux is 
\begin{equation}
F_c=-\kappa_0\, T^{5/2}\, \frac{dT}{dS}
\end{equation}
with $T$ temperature, $\kappa_0$ thermal conductivity coefficient, taken as
the standard value $\kappa_{\rm Sp} = 10^{-6}\rm{g cm/(sec^3\bf{(K)}^{7/2})}$ (Spitzer
1962). Heat conduction is an important feature in the model,
different to a cooling flow model. We evaluate the structure for
different values of the thermal conductivity $\kappa_0$, 1/5, 1/25 and 1/100 of 
$\kappa_{\rm Sp}$. 
Mass conservation gives
\begin{equation}
\dot M=2\pi r^2 \rho v =const.
\end{equation}
with $\dot M$ accretion rate, $\rho$ density and  $v$ flow 
velocity (positive if downward directed). Euler's equation in the
stationary 
case is  
\begin{equation}
 v\frac{dv}{dS} +\frac{1}{\rho}\frac{dP}{dS} - \frac{GM_*}{r^2}=0
\end{equation}
with $P$ pressure, $G$ gravitational constant, $M_*$ mass of the white
dwarf.  Energy conservation gives 
\begin{eqnarray}
\frac{1}{r^2}\frac{d}{dS}\:\left[\frac{\dot
M}{2\pi}\left(\frac{1}{2}v^2+ \frac{GM_*}{r_*}-
\frac{GM_*}{r} +\frac{\gamma}{\gamma-1}\frac{P}{\rho}\right)+r^2F_c\right]\\
=-\Lambda(T)\: n_e\, n   \nonumber
\end{eqnarray} 

with $\gamma=\frac{5}{3}$ the ratio of specific heats for the fully
ionized gas, atom and electron densities $n$, $n_e$ and $\Lambda(T)$
the radiative loss function for solar abundance 
(Sutherland \& Dopita 1993). In our highly subsonic solutions the
kinetic energy term is negligibly small.We have normalized the gravitational
potential energy to zero at the white dwarf surface.
(Following Shmeleva $\&$ Syrovatskii (1973) we had added a
source term to account for the finite surface temperature of the white
dwarf, see Liu et al. (1995)).
We solve these equations together with the equation of state.

We assume that before radiative losses of the settling gas become
important, the accretion flow has already lost its angular momentum, with
which it started from the corona at the disk truncation radius, and
that the rotational energy has been dissipated into heat. The maximum
energy which this flow can bring with it is then the gravitational
energy between the disk truncation radius and the white dwarf surface
where the former term can be neglected if the truncation radius is
many white dwarf radii away. Support for our neglect comes from the
 boundary layer rotation velocities derived by Pandel et al. (2005)
for the dwarf novae in their sample, found to be considerably smaller 
than the Keplerian velocity near the white dwarf. 

Our boundary conditions are the following.
The lower boundary is located at the white dwarf surface. The temperature is
taken equal to the black body temperature arising from the irradiation
of the surface by the downward directed half of the radiative energy
release.
For the location of the upper boundary we take a height above the white
dwarf surface above which the radiative loss is negligible. At this height we
require that the total energy flow, i.e. the sum of advective,
gravitational and thermal conductive flow, equals the total
available accretion energy flow. These energies are
\begin{eqnarray}
\dot E_{\rm{conductive}}&=&2\pi r^2 F_c \\
\dot E_{\rm{advective}}&=&\dot M\, \frac{\gamma}{\gamma -1}
\frac{\Re}{\mu} T \\
\dot E_{\rm{gravitational}}&=&\dot M \,\frac{GM}{r_*}\left(\frac{1}
{r_*}- \frac{1}{r}\right)
\end{eqnarray}
with $\Re$ gas constant and $\mu$ molecular weight (taken as 0.6).

This boundary condition neglects outward conductive heat 
losses at large radii where the temperature drops outwards. Such
losses have been estimated to be small in Liu et al. (1995) where they
amount, for full Spitzer conductivity, to the order of magnitude of
the evaporation energies at large radii. (Real temperatures would come
out slightly lower than with this neglect.) After taking the mass
accretion rate that corresponds to the observed flux, our model
contains no free parameter except the assumed value of thermal
conductivity, which we want to test.

\section{Comparison of different models}
In an isobaric cooling flow (Mushotzky \& Szymkowiak 1988) the
emission measure for each temperature decrement is determined by
the time it takes the matter to radiatively cool down to the next
temperature shell. No thermal conductivity is 
included. To simulate the observed spectrum a mass flow rate that
corresponds to the observed flux is taken, in addition as a free
parameter a maximal temperature is chosen to fit the observations.

Narayan and Popham (1993) investigated the structure of thin
accretion disks around a central white dwarf with emphasis on a
self-consistent description of the boundary layer. The decrease of
rotation from
the Keplerian value at the upper boundary to the value at the stellar
surface as well as friction is taken into account. For low rates as
appropriate for the dwarf nova systems in quiescence they found an
optically thin very hot boundary layer. Heat is 
transported advectivily into the optically thick layers of the white dwarf from
where it is radiated as X-rays. These investigations are carried out only
for accretion rates 10 times higher or even more than in our VW Hyi 
analysis. 

The solutions based on a hot settling flow by Medvedev \& Narayan
(2001) and Medvedev \& Menou (2002) also include the rotational energy of
the accreting gas, so that the white dwarf rotation and the viscous
nature of the flow are accounted for. 
Mahasena \& Osaki (1999) model the boundary layer structure as
determined by thermal conduction in the settling gas around the white
dwarf, similar to the 'siphon-flow model' for a disk corona 
(Meyer \& Meyer-Hofmeister 1994), and also similar to our white dwarf
boundary modeling.

\section{Results of computations}
\subsection{Energy flows in the boundary layer}

We solve the system of differential
equations given in the previous section, the same procedure as carried
out for the earlier investigation (Liu et al.1995).
For our comparison of theoretical and observed spectra for VW Hyi we 
take $0.62 M_\odot$ for the white dwarf mass and $8.3\cdot 10^8$ cm
for its radius. We do not perform computations for different
abundances since the aim of our analysis is the study of the influence
of thermal conductivity. The change of abundances allows better fits
of the observed spectrum. For a discussion of the abundances present 
in VW Hyi see the discussion in Pandel et al. (2005).
The free parameter in our model is the mass accretion rate. We take $\dot
M=1.25\: 10^{-12}M_\odot/yr$ assuming that the matter is accreted
on an equatorial belt covering half of the white dwarf surface area. 
Because we find that
100\% the Spitzer value does not give an acceptable agreement with the 
observations, we vary the value of heat conductivity. In the following we
show results for 1/100 and 1/25 and 1/5 of the standard Spitzer value.

\begin{figure}
\includegraphics[width=7.8cm]{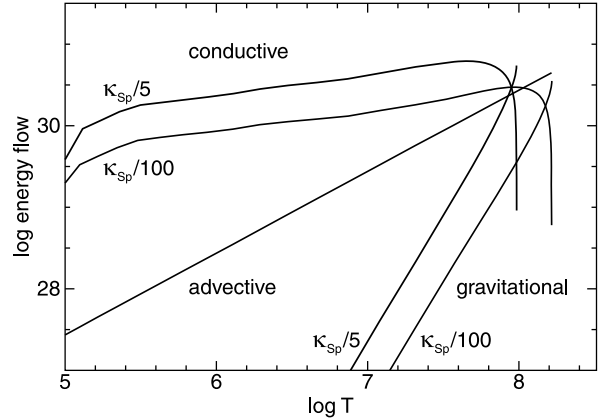}
\caption{Energy flux versus temperature for heat conductivity
 1/5 and 1/100 of the Spitzer value. Note the higher temperature
maximum, $1.6\cdot 10^8$ K, for the lower conductivity value}
\label{energy5-100}
\end{figure}

In Fig. \ref{energy5-100} we show the 
evaluated change of energy flows with increasing height, that is
increasing temperature for a heat conduction reduced to 1/5 and 1/100 
of the standard value. The upper
boundary location (according to the requirements discussed in the last
section) is at different heights above the white dwarf surface,
at $6.8 \cdot 10^8$ cm in the first case and $1.8 \cdot 10^9$ cm  in the
second case. For the lower conductivity the advective energy is
dominant at high temperature. The most important difference is the
maximum temperature reached.
This has a strong influence on the resulting emission measures (and
the spectrum) as will be shown later. In our model the maximum
temperature is a result of the structure computations. Note that in the
cooling flow model the maximal temperature is chosen to fit the spectrum.

\subsection{Emission measures}
In Fig.\ref{emission} differential emission measures (emission measures
per $\rm{cm^2}$ surface area, per degree) [$\rm{cm^{-5} K^{-1}}$] are 
shown for a heat
conductivity 1/5 and 1/100 of the standard 
Spitzer value. For the higher value of heat conductivity the
differential emission measures are systematically higher, but with no
contribution at high temperature.

\begin{figure}
 \includegraphics[width=7.8cm]{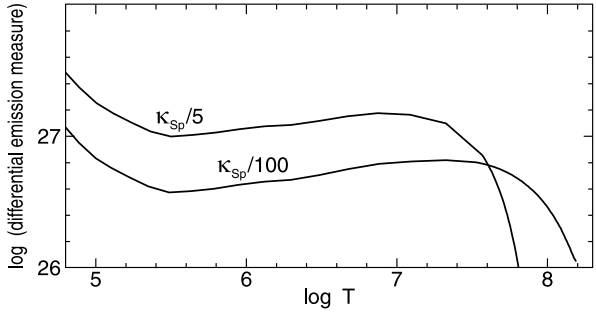}
 \caption{Differential emission measures per surface area 
[$\rm{cm^{-5} K^{-1}}$] for heat conductivity 1/5 and 1/100 of the standard 
Spitzer value. }
\label{emission}
\end{figure}

To obtain the radiation from the boundary layer, the differential emission
measures have to be multiplied by the area from which
the radiation comes, the temperature $T$ and the cooling function 
$\Lambda(T)$. The surface area is the spherical shell around the white
dwarf at the height with temperature $T$. Since all the boundary layer
is close to the white dwarf this flaring effect in the geometry only
occurs at the highest temperatures. This product, the evaluated
contributions to the luminosity from different temperature regions, is
shown in Fig.\ref{lum}. The departure from smooth lines is due to the
non-smooth dependence of the cooling function on temperature. With
this display of luminosity
distributions it becomes clear that different contributions arise
from the structure at high temperatures. And, as shown in the next
section, agreement with the
observed spectrum for VW Hyi can  only be gained with the high
temperature contributions which result from the low conductivity.

\begin{figure}
\includegraphics[width=7.8cm]{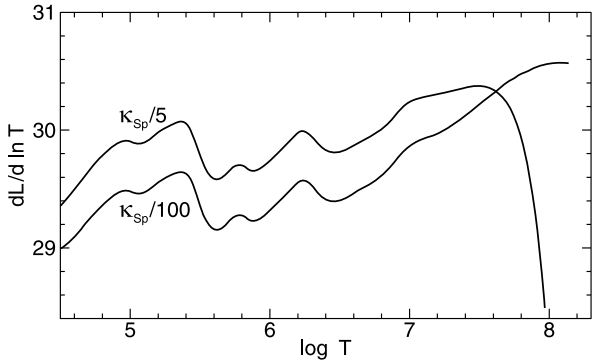}
 \caption{Total emission per logarithmic temperature interval for heat 
conductivity 1/5 and 1/100 of the standard Spitzer value. }
\label{lum}
\end{figure}

The emission measures of a cooling flow model are (see Pandel et al. 2005,
taken from Mushotzky \& Szymkowiak 1988) 
\begin{equation}
\frac{dEM}{dT}=\frac{5}{2}\frac{\Re}{\mu} \dot M \frac{n^2}
{\varepsilon(T,n)}
\end{equation}
with $\varepsilon(T,n)$ the total emissivity per volume. The total
emission per logarithmic temperature then becomes
$dEM/dT \:\Lambda(T)\, T = \frac{5}{2}\frac{\Re}{\mu}
\dot M T$. Pandel et al. (2005) modified the cooling flow model 
emission measures adding a factor
$\left(T /T_{\rm{max}}\right)^\alpha$, with $\alpha$=-0.05 for VW Hyi to
improve the fit of the spectrum, and derived an accretion rate of $3.7\cdot
10^{-12}\: \rm{M_\odot/yr}$. We use $1.25\cdot 10^{-12}\:
\rm{M_\odot/yr}$ for our calculation, the matter accreting onto half
the white dwarf surface. 
To compare the cooling flow radiation with our results for low
conductivity we have multiplied these cooling flow emission measures 
(without $\alpha$ modification) by a factor of 0.676 (=ratio of
accretion rates, our rate doubled) to account for the different
accretion rates and accretion areas. The result is shown in
Fig.\ref{em_cf}. The
contributions from the cooling flow model lie above those from our
model, but end at the maximum temperature
chosen for the fit, $T_{\rm{max}}=9.5 \cdot 10^7$ K, so the emission is
roughly compensated.

\begin{figure}
\includegraphics[width=7.8cm]{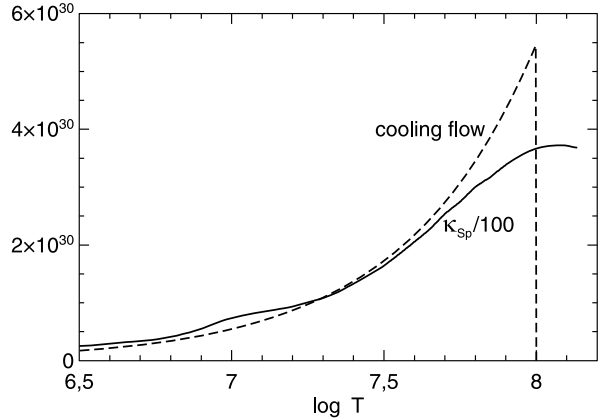}
 \caption {Comparison of total emission per logarithmic temperature interval
for a cooling flow model and our model for low conductivity}
\label{em_cf}
\end{figure}

\section{Spectra}
We use the X-ray spectral fitting package XSPEC (version 11).
We evaluated the spectra for the three sets of boundary layer
structure computed for the conductivity 1/5, 1/25 and 1/100 of
the standard Spitzer value. We show these spectra in 
Fig. \ref{spectrum} (a), (b), and (c). The
spectra document how the thermal conductivity determines the
distribution of radiation over the energy range. 
A comparison of Fig. \ref{lum} shows how the different 
contributions to the radiation for high and low conductivity values 
determine the spectrum for temperatures around $10^7$ and $10^8$.
Below each spectrum the $\chi$-statistics values are shown. Note the
different scales in Fig. \ref{spectrum} (a), (b), and (c). Only for
the low conductivity do we find an acceptable agreement.

\begin{figure}
\includegraphics[width=9.4cm]{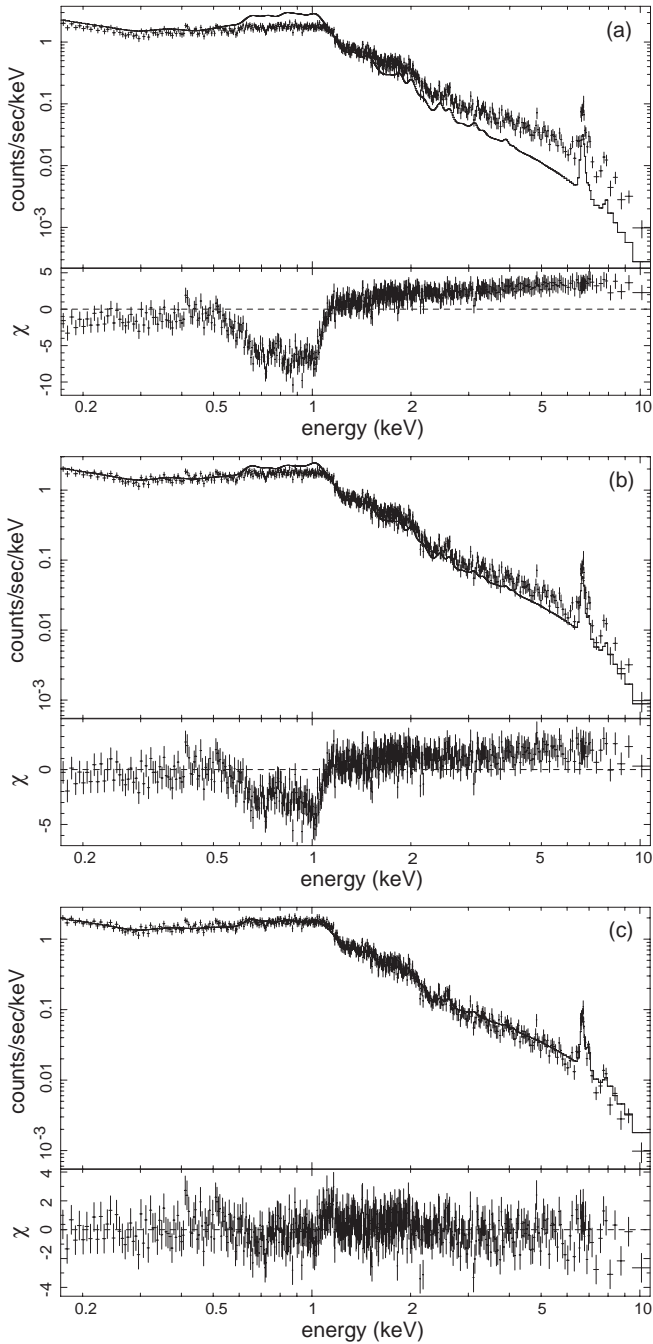}
 \caption {Comparison of XMM-Newton EPIC spectra for VW Hyi in
 quiescence with model spectra evaluated for an accretion rate of $\dot
 M=1.25\cdot 10^{-12}\, \rm{M_\odot/yr}$  and for different 
thermal conductivity:   (a) $1/5 \kappa_{\rm{Sp}}$, (b) $1/25
 \kappa_{\rm{Sp}}$ and (c) $1/100 \kappa_{\rm{Sp}}$.}
\label{spectrum}
\end{figure}

We have not tried to obtain the best fit for a variation of the mass accretion
rate, nor for changes of the abundances, and we have not analyzed
spectral lines, as was done in previous work (Pandel et al. 2003,
Perna et al. 2003, Pandel et al. 2005). The aim of our investigation
is to study the influence of the heat conductivity.

We compare the accretion rate roughly found from our modeling, 
$1.25 \cdot 10^{-12}\,
\rm{M_\odot/yr}$, with the rates derived 
in other investigations. Pandel et al. (2003) found $5 \cdot 10^{-12}\,
\rm{M_\odot/yr}$ and Pandel et al. (2005)
$3.7 \cdot 10^{-12}\,\rm{M_\odot/yr}$, for fits to the same
data. Previous observations for VW Hyi lead to an estimate of $3 \cdot
10^{-12}\,\rm{M_\odot/yr}$ for the accretion rate in late quiescence 
(BeppoSAX observations by Hartmann et al. 1999); similar results were
gained from earlier observations (EXOSAT and ROSAT observations by van 
der Woerd et al. 1987, Belloni et al. 1991). From theoretical arguments one
would expect that the X-ray flux changes during the quiescent interval
in the outburst cycle. One would expect an increase of the rate in
quiescence due to the increasing amount of matter accumulated in the
disk and the increasing mass flow rate (Meyer-Hofmeister \& Meyer 1988).

We note that the observations  also include the radiation from the
accretion disk. In earlier work we compared the amount of X-rays from
the accretion disk and from the white dwarf boundary layer for VW Hyi 
(Meyer et al. 1996, Fig. 1) and found that the distribution of
radiation over the energies is similar in both cases, but the total 
contribution from the boundary layer is much smaller. This means that the
shape of the spectrum of the white dwarf boundary layer should not be 
essentially changed by an 
additional contribution of disk X-ray emission and we can interpret
the observed spectrum as the spectrum of the boundary layer. 

\section{Discussion}
\subsection{Spectra from different models}
The first comparisons with emission line spectra from the new X-ray 
telescopes were carried out for other dwarf novae, not VW Hyi. Ramsay et
al. (2001b) presented a spectrum of OY Car (in quiescence) obtained
 during the performance phase of the  \it{XMM-Newton} 
\rm {mission (Ramsay et
al. 2001a). They found that the spectrum was best fitted by a three 
temperature MEKAL thermal plasma model with a partial covering
absorber. Mukai et al. (2003) discussed }\it {Chandra }\rm {HETG spectra
of seven CVs and found for the non-magnetic systems EX Hya, V603 Aql,
U Gem and SS Cyg that they are remarkably well fit by a cooling flow model.

Perna et al. (2003) performed the first comparison with theoretical spectra,
evaluated for the models by different authors, which were available at that
time. Their very detailed analysis for WX Hyi in quiescence includes
line effects, not included in our analysis.
The results were : (1) For the cooling flow model they found it
accounts reasonably well for the continuum of WX Hyi and the line
strength in the short-wavelength region, but under-predicts the
emission at long wavelength. (2) For the model of Narayan \& Popham (1993) 
they found that the increase of density in the outer regions results
in substantially more emission in O VII and O VIII lines than in a
cooling flow. The agreement between observed and predicted low
temperature lines 
is reasonable and could be due to the assumed bremsstrahlung
cooling. (3) For the coronal siphon flow model of Meyer \&
Meyer-Hofmeister (1994), emission measures derived for the
transition layer between corona and disk were taken. However the
more important emission measures for the accretion layer on the white
dwarf were given in Liu et al. (1995). 
(4) The spectra from the X-ray emitting corona model of Mahasena \&
Osaki (1999), for the assumed low accretion rate, plausible for WX Hyi,
show a too soft continuum and too strong emission lines. But the fit
with higher accretion rate for U Gem was better. (5) For the hot
settling flow solutions (Narayan \& Medvedev, Medvedev \& Menou 2002) 
the accretion rate was
adjusted to a value to reproduce the observed X-ray luminosity, but
the X-ray line emission seems greatly over-predicted.

For VW Hyi a first spectral analysis was performed by Pandel et
al. (2003). They found the best agreement
between observations and computed spectrum for a CEMEKL model,
slightly better than for a three-temperature model, and also a good fit for
the MKCFLOW cooling flow model. The authors pointed out a
remarkable qualitative agreement with the model of 
Narayan \& Popham (1993). In the spectral analysis of Pandel et al. (2005)
based on }\it{XMM-Newton} \rm {data, including VW Hyi, it was found
that, in general, the X-ray emission originates from a hot,
optically thin multi-temperature plasma with a temperature distribution
in close agreement with an isobaric cooling flow.

These discussions of spectra resulting from different models only
allow the conclusion that spectra of a cooling flow model or a
combination of a few MEKAL one-temperature models give a good fit. To 
really test the other boundary layer models, 
detailed calculations for a chosen dwarf nova system would have to be
carried out (as done for our model) to compare the model with the observed spectrum.

\subsection{Spectra of other dwarf novae} 
The investigation of ten dwarf novae by Pandel et al. (2005) allows a
comparison of the results for the different systems.
From Fig.1 in their work showing all spectra together, the spectra of the eight dwarf novae in quiescence look
qualitatively similar, except that of OY Car (the lower flux at low
energies attributed to intrinsic absorption) and EI UMa (the harder
spectral slope (interpreted as characteristic for the intermediate
polar objects).  An important result is the maximal
temperature found for these systems. For seven of these systems (U Gem
omitted) the maximal temperatures found from the
fits lie in the range from $9.5\, 10^7$K (VW Hyi) to $3\,10^8$K (WX Hyi).
That means to fit the spectra of the other systems we also would have
to assume a low heat conductivity.

\subsection{High and low thermal conductivity}

There is evidence for both high and low thermal conductivity in different 
astrophysical situations. 

On one hand, observations indicate that thermal conduction across the
temperature jumps of cold fronts in clusters of galaxies
(e.g. Abell 2142 and Abell 3667) is far below the Spitzer value 
(Markevitch et al. 2000, Ettori \& Fabian 2000, Vikhlinin et
al. 2001). This is interpreted as resulting from 
gas motion on both sides of the interface that produces a magnetic
field stretched parallel to the interface (Vikhlinin et
al. 2001). SPH calculations relate the width of the
cold front to the value of the thermal conductivity and support this
explanation, yielding a reduction to the percent level of the Spitzer value
(Asai et al. 2004, 2007, Xiang et al. 2007).

On the other hand, modeling the temperature gradients of cooling
flows in clusters of galaxies Zakamska \& 
Narayan (2003), Ghizzardi et al. (2004), and Voigt \& Fabian (2004)
found evidence for thermal conduction of about 30 up to near 100\%
of the Spitzer value. Narayan \& Medvedev (2001) suggested that such 
high values can result in tangled and chaotic magnetic
fields (Rechester \& Rosenbluth 1978, Chandran
\& Cowley 1998) if the field is chaotic over a wide 
range of length scales
(factors of 100 or more) as might happen in strong MHD turbulence
(Goldreich \& Sridbar 1995). In contrast, low thermal conductivity
should result for weak MHD turbulence (effectively a
superposition of Alfv${\acute{e}}$n waves).

High thermal conductivity, near the Spitzer value, appears 
also to be present in coronal boundary layers above accretion
disks. In X-ray binaries, features like spectral state 
transitions and hysteresis can be well modeled in 
the framework of disk evaporation (Meyer \& Meyer-Hofmeister 1994, 
Meyer et al. 2000) that takes the exchange of mass and energy 
between disk and corona into account. In such models the flow of 
mass from the disk into the corona depends significantly on the 
value of the thermal conductivity (Meyer-Hofmeister \& Meyer 2006). 
Quantitative agreement with observations results for high 
values of the thermal conductivity.

\subsection{High conductivity a result of disk dynamo action?} 

The high conductivity found in the cases discussed above differs 
remarkably from the low conductivity found here, for an 
apparently similar accretion process, here from a hot 
corona to the cool white dwarf surface. Narayan \& 
Medvedev's (2001) suggestion implies strong MHD turbulence 
in the layers between corona and disk but not in the corresponding 
layers between corona and white dwarf surface. Stable layering 
of the cooling gas in the strong gravity of the white dwarf
might suppress turbulence in the white dwarf case.

High thermal conductivity in the coronal transition layers of 
accretion disks could also result from magnetic fields generated 
through dynamo action in the accretion disks. These fields 
attain energy densities that are a fraction of the disk internal 
pressure but reach out into the low density atmosphere of the 
disk where they become dominant over the gas pressure (Hirose et
al. 2006). One also notes that magnetic diffusion 
in the accretion disk transports magnetic power from small scales, 
of the order of the disk scale height (on which the fields 
are generated) to large scales, up to  the size of the local disk 
radius (Brandenburg et al. 1995). Large 
scale unipolar fields reach farther than a few scale 
heights of the disk into the corona where they will merge with coronal
magnetic fields. Continuous reconnection could result in a 
fairly direct thermal path along magnetic field lines from corona 
to disk and provide high thermal conductivity in the 
conductively heated and radiatively cooling transition layer.

\subsection{Magnetic fields advected in white dwarf accretion}

The very low thermal conductivity in white dwarf accretion 
raises an important question. The high thermal 
conductivity \it{along} \rm{ magnetic field lines suggests thermal 
insulation in these accreting layers by a nearly horizontal magnetic 
field. Is this a natural outcome of accretion from a hot 
ADAF-like coronal flow with turbulent magnetic fields? We show that this 
might point to a very different accretion geometry than that 
commonly considered.

The mean magnetic energy density in accretion disks obtained in 
magneto-hydrodynamic simulations typically is a fraction of 
the gas energy density, e.g. one quarter (Hirose et
al. 2006, Sharma et al. 2006, the latter for collision-free plasma). 
Due to shearing Kepler motion 
the azimuthal component of the mean magnetic field is 
larger that the poloidal components, e.g. by a factor 
of 6. Treating the ADAF as a thick accretion disk, we now can 
estimate the strength of the azimuthal magnetic field in the  
gas approaching the white dwarf. This gas and 
its magnetic field become highly compressed as they settle in the 
strong gravity on the white dwarf surface. The compression ratio can be 
estimated by comparing the density in the  ADAF near the 
white dwarf with the density in the cooling layer. Using 
the cooling layer solution for VW Hyi and the ADAF solution 
of Narayan et al. (1998) (for $\alpha$=0.3 and the same mass flow 
rate), one obtains a factor $10^{1.9}$ by which the density is increased.

With flux conservation, the 
toroidal component of the magnetic field increases by 
the same factor while the radial component only increases by 
a geometrical factor, the ratio of the ADAF surface to the 
white dwarf surface,  a factor of 4 if the ADAF ends at 
a height of one stellar radius above the white dwarf surface. 
Since we are interested only in an approximate order of 
magnitude the exact numbers are not important. Also, the 
possible weakening of the toroidal flux by compression of 
oppositely directed field lines and annihilation may be neglected. 
(The mass that ends up in the settling layer has previously
occupied an ADAF region of the size of a radius, the 
typical size of magnetic fields created by a coronal dynamo.) 

This simple exercise yields a ratio between the magnetic field 
components parallel and perpendicular  
to the surface of the white dwarf of the order of 100, that 
would result in a corresponding reduction of the radial thermal 
conductivity by the same factor. However, the same estimate 
yields a magnetic field strength whose pressure far exceeds the 
gas pressure in the cooling region, by about a factor of 100. 
This invalidates the assumption of a plane gas pressure supported 
cooling flow, with interesting consequences.

\subsection{Cooling gas suspended in magnetic fields?}

Cooling gas supported against gravity by gas pressure and the magnetic 
pressure of a horizontal magnetic field becomes strongly unstable 
to the Parker instability when the magnetic pressure becomes a
significant part of the total pressure: an alternating vertical 
up-and-down displacement of an 
initially horizontal flux tube creates peaks and troughs so that 
matter can flow along the flux tube to collect in the troughs 
and evacuate the peaks. The troughs become heavier than their 
surroundings and sink down, the peaks become buoyant and rise. If 
the magnetic tensions become dominant, as in our case, this will 
end up in narrow filaments of matter magnetically suspended above 
the white dwarf surface, reminiscent of solar filaments of cool 
chromospheric gas magnetically suspended in the corona. The transport 
of angular momentum of the accreting gas, of course, needs further 
consideration. We only note that the magnetic field configuration
might lend itself to effective removal of angular momentum.

Thus from the simple advection of coronal/ADAF magnetic fields
a spatially intermittent accretion flow will arise, possibly 
involving the formation of accretion spots. It is interesting 
that this very different geometry preserves the thermal 
insulation of the cooling gas: gas of different temperature 
resides in different flux tubes. Would this situation also lead 
to cooling under constant pressure as the analysis of observed 
spectra indicates? The gas pressure in the flux tubes is the 
weight of the column density, and as the column density remains
constant on cooling, so does the pressure. Thus our 
good fits to the observed spectra could be the natural signature 
of accretion from a hot magnetized ADAF on non-magnetic white 
dwarfs in the quiescent stage of dwarf novae cycles.

\section{Conclusions}

We have calculated the structure of the 
boundary layer that forms at the surface of non-magnetic 
white dwarfs on accretion from a hot corona in quiescent 
dwarf nova systems. Our calculated spectra significantly 
depend on the value of the thermal conductivity. Comparing 
our results with observed spectra for VW Hydri we find 
good agreement for very low values of the conductivity, 
of the order of 1\% of the Spitzer value for ionized gases. 
For higher values too much energy is drained from the 
hottest layers and radiated at cooler temperatures to 
give a satisfactory fit.

This suggests thermal insulation of layers of different 
temperature from each other by magnetic fields. A 
discussion of high and low conductivity 
cases in various circumstances  leads us to the 
suggestion of spatially intermittent accretion in 
magnetic fields, preserving the low thermal conductivity 
isobaric cooling that our spectral fits require.

\begin{acknowledgements}
F.K. Liu acknowledges the support of the 
National Natural Science Foundation of China (No. 10573001) and thanks 
MPA for hospitality.
\end{acknowledgements}

{}

\end{document}